\documentstyle[12pt]{article}
\newcommand{\bm}[1]{\mbox{\boldmath$#1\!$}}
\newcommand{\bn}[1]{\mbox{\boldmath$#1$}}

\newcommand{\beq}{\begin{equation}}
\newcommand{\eeq}{\end {equation}}
\newcommand{\bea}{\begin{eqnarray}}
\newcommand{\eea}{\end{eqnarray}}

\begin{document}
\lineskip=24pt
\baselineskip=24pt
\raggedbottom
{\Large{\centerline{\bf Electrodynamics of Bose-Einstein condensates }}}
{\Large{\centerline{\bf in angular motion}}}
\vskip 1.0cm
\centerline{{\bf  L. G. Boussiakou, C. R. Bennett} and {\bf M.
Babiker}}
\centerline{ Department of Physics, University of York, Heslington,
York YO10
5DD, England}
\vskip 1.0cm
\section*{Abstract}

A theory determining the electric and magnetic properties of vortex states in
Bose-Einstein condensates (BECs) is presented.  The principal ingredient is the Lagrangian
of the system which we derive correct to the first order in the atomic centre of mass
velocity. For the first time using centre of mass coordinates, a gauge transformation is performed and relevant relativistic corrections are included.
The Lagrangian is symmetric in the electric and magnetic aspects of the problem and
includes two key interaction terms, namely the Aharanov-Casher
and the R\"{o}ntgen interaction terms.  The constitutive relations, which link the
electromagnetic fields to the matter fields via their electric polarisation and
magnetisation, follow from the Lagrangian as well as the corresponding Hamiltonian.
These relations, together with a generalised Gross-Pitaevskii equation, determine
the magnetic (electric) monopole charge  distributions accompanying an order n vortex state 
when the constituent atoms are characterised by an electric dipole (magnetic dipole). Field distributions
associated with electric dipole active (magnetic dipole active) BECs in a vortex state are evaluated for an 
infinite- and a finite-length cylindrical BEC. The predictd monopole charge 
distributions, both electric and magnetic, automatically satisfy the  requirement of global charge neutrality
and the derivations highlight the exact symmetry between the electric and magnetic properties.  Order
of magnitude estimates of the effects are
given for an atomic gas BEC, superfluid helium and a spin-polarised hydrogen BEC.
 \newpage
 \section{Introduction}
 
Recent years have seen the emergence of the physics of cold atomic
ensembles and Bose-Einstein condensates (BECs), thanks to advances in laser cooling and 
trapping [1]. Atomic ensembles can now be routinely cooled and confined and, ultimately, made to condense under 
various conditions [1-3],
most notably in magnetic and light traps,  near surfaces, in material cavities and in optical lattices.
The trend is towards producing larger systems with increasing numbers of constituents 
and increasing densities than currently.  Furthermore, continuous evaporative cooling in optically trapped 
bosonic and fermionic gases have been demonstrated in stable optical traps and the techniques 
promise the attainment of BECs with very high densities and much larger coherence lifetimes [4].

In parallel with these experimental advances, various theoretical investigations [5]
have been made,  with increasing recent emphasis 
directed towards the angular motion of BECs in the form of vortex states [6-9] and the 
electromagnetic features associated with them. In particular,
Leonhardt and Piwnicki [7] examined the magnetic effects due to the angular motion of a  BEC whose constituent atoms
are characterised by electric dipoles, thus forming a quantum dielectric in angular motion.  They predicted that such a motion 
should give rise to a magnetic monopole positioned at the core of an $n=1$
vortex. More recently, Shevshenko considered the converse case and put forward a theory for the generation of vortices in 
superfluid films by application
of crossed electric and magnetic fields [9]. 

The situation is now ripe for a rigorous theory of the electromagnetic properties of BECs in 
rotational motion to be constructed.  In contemplating such a theory, an immediate question which springs to mind is as follows.
Hitherto, only the electromagnetic properties of dielectric BECs, namely those whose constituent atoms are characterised by an electric dipole moment (electric BECs),
have been considered.   What if the BEC atoms are characterised by a magnetic
dipole moment (magnetic BECs)?  This question is now of practical relevance since BECs can be created in light
traps, allowing, the magnetic properties of, for instance, alkali metal BECs
to be investigated [8].  More significantly, explicitly magnetic BECs,  particularly spin-polarised hydrogen BECs, have 
been created [10]. 
Intuitively one expects the physics of the problem to display a symmetry between the electric
and magnetic aspects.  Thus if a rotating electric BEC generates a magnetic monopole,  then a rotating magnetic BEC 
should generate an electric monopole. One of the main tasks of the theory presented here is to test the validity of such an expectation. 

In this paper we show that rotating magnetic BECs are indeed predicted to give rise to electric monopoles and that
 it is possible to formulate a theory in which
both types of effect arise in a unified manner.  However,  our theory predicts that rotating BECs should not, as in Ref.[7] give rise to
localised monopoles,  but, rather, to monopole
distributions.  The treatment automatically ensures that charge neutrality is correctly preserved by
the electric and magnetic monopole distributions associated with the  
vortex states of the two types of BEC. 

The outline of this paper is as follows.  In section 2 we begin by introducing the model system as  
a single atom interacting with the electromagnetic potentials via the familiar charge and current sources.
We then systematise
the procedure based on a gauge transformation leading from the conventional Lagrangian of this system to a form displaying the 
convective velocity-dependent interactions.  These interactions involve the coupling of the electric and magnetic field intensities
to the entire series of electric and magnetic mutipoles in exact closed forms.  In particular, we identify the R\"{o}ntgen interaction [11]
 and  we make use of symmetry arguments to incorporate an Aharanov-Casher interaction term [12].  In section 3 we derive the 
 field equations which permit the identification of constitutive relations connecting matter fields to electromagnetic fields.  
We also derive the single-atom plus fields
Hamiltonian  and make use of this in section 4 to construct the many-body formulation appropriate for an ensemble of such atoms 
forming a BEC.  A main outcome of the many-body theory is a systematic derivation of a generalised Gross-Pitaevskii equation.
In section 5 we consider rotating BECs and solve the constitutive relations, together with the Gross- Pitaevskii equation,
to determine distributions for an order n vortex state of the two types of BEC. We also display
the results for typical situations involving order of magnitude estimates of the effects,
evaluated for an atomic gas BEC, superfluid helium and spin-polarised hydrogen BEC.  Section 6 concludes with  a summary
and provides further comments.  

\section{Lagrangian}
It suffices to model the constituent BEC atom in terms of a neutral
two-particle
system of mass $M=m_{1}+m_{2}$ involving two bound charges 
$e_{1}=-e_{2}=e $ of masses $m_{1}$ and $m_{2}$, position vectors 
${\bf q}_{i}$ and velocities ${\bf {\dot q}}_{i}$ where $i=1,2$.  The
conventional Lagrangian of  
this two-particle system coupled to the electromagnetic scalar potential $\phi$
and vector potential  ${\bf A}$ in an arbitrary gauge 
is written in the following form, which is non-relativistic 
in the motion of both particles [14-17],
\beq
 L=\frac{1}{2}m_{1} \dot{q}^2_1 + \frac{1}{2}m_2 \dot{q}^2_2 + \int d^3{\bf r} \left[
\frac{1}{2}\epsilon_0 \left({\bf E}^2({\bf r}) - c^2 {\bf B}^2({\bf r})\right)
+{\bf J}({\bf r}).{\bf A}({\bf r})-\rho({\bf r})\phi({\bf r}) \right],
\label{1}
\eeq
where ${\bf E}=-\dot{\bf A}-\nabla \phi$ and ${\bf B}=\nabla\times{\bf
A}$. 
The charge and current densities are those appropriate for point charges, namely
\beq
\rho({\bf
r})=\sum_{i=1,2}e_{i}\delta({\bf r}-{\bf q}_{i});\;\;\;\;\;\;
{\bf J}({\bf r})=\sum_{i=1,2}e_{i}{\bf {\dot q}}_{i}\delta({\bf r}-{\bf q}_{i})\label{2}
\eeq
In order to reveal the bound nature of the two particles and exhibit the desired effects involving gross motion of the whole 
atom we need to express the theory in  terms of the centre of mass and relative coordinates.  We therefore introduce the centre of 
mass coordinate ${\bf R}$ and internal coordinate ${\bf q}$ by [16]
\beq
{\bf R}=\frac{(m_1 {\bf q}_1 + m_2 {\bf q}_2)}{M};\;\;\;\;\;\; {\bf q}={\bf q}_1 - {\bf q}_2\label{3}
\eeq
This step paves the way for the application of a Power-Zienau-Woolley (PZW) gauge transformation [16], 
to be carried out next at the Lagrangian level. As far as the authors are
aware, this is the first time this method has been performed when using centre
of mass coordinates, although similar results were obtained using a
multi-polar expansion on the equivalent Hamiltonian [14]. The generator
of the gauge transformation is defined by
\beq
\chi({\bf r})=\frac{1}{e}\int d^{3}{\bf r'}{\bf A}({\bf r'}).{\bm {\cal F}}_{m}({\bf r,r'})
\eeq
where ${\bm {\cal F}}_{m}({\bf r,r'})$ is a gauge vector function given by
\beq
{\bm {\cal F}}_{m}({\bf r,r'})=e\int_{0}^{1}d\lambda({\bf r-R})\delta({\bf r'-R}-\lambda({\bf r-R)})
\eeq
Note that despite superficial appearances, this vector function is not the electric multipolar polarisation 
field in closed form. The gauge transformation involves replacing the vector potential ${\bf A}$ and scalar potential $\phi$ in the
Lagrangian Eq.(\ref{1}) by ${\bf A'}$ and $\phi'$ related to the untransformd potentials by
\beq
{\bf A'}({\bf r})={\bf A}({\bf r})-{\bn {\nabla}}\chi(r);\;\;\;\;\;\;\;\;\;
\phi'({\bf r})=\phi({\bf r})+\frac{\partial\chi({\bf r})}{\partial t}
\eeq
The transformed Lagrangian $L'$ is
\bea
L'=L
-\int d^{3}{\bf r}\left\{{\bf J}({\bf r}).{\bn {\nabla}}\chi({\bf r})+\rho({\bf r})\frac{\partial\chi({\bf r})}{\partial t} \right\}
\eea 
It is straightforward, albeit laborious, to show that the terms involving the gradient and time derivative of the gauge function
are given by
\beq
{\bm {\nabla}}\chi={\bf A}({\bf r})+
\frac{1}{e}\int d^{3}{\bf r'}\;{\bm {\Theta}}_{m}({\bf r,r'})\times
{\bf B}({\bf r'})
\eeq
and 
\bea
\frac{\partial\chi}{\partial t}&=&\frac{1}{e}\int d^{3}{\bf r'}\; 
\left\{{\bf {\dot A}}({\bf r'}).{\bm {\cal F}}_{m}({\bf r,r'})+
\left({\bf {\dot R}}.\frac{\partial}{\partial{\bf R}}\right)
\left[{\bf A}({\bf r'}).{\bm {\cal F}}_{m}({\bf r,r'})\right]\right\}
\nonumber \\
&=&-{\bf {\dot R}}.{\bf A}({\bf R})+\frac{1}{e}\int d^3{\bf r'}\;
\left\{{\bf {\dot A}}({\bf r'}).{\bm {\cal F}}_{m}({\bf r,r'})+
{\bf {\dot R}}.\left[\left({\bm {\cal F}}_m({\bf r,r'})-{\bm \Theta}
_m({\bf r,r'})\right)\times{\bf B}({\bf r'})\right] \right\}
\nonumber \\
\eea
with
\beq
{\bm \Theta}_m({\bf r,r'})=e\int_{0}^{1}d\lambda\lambda({\bf r-R})
\delta({\bf r'-R}-\lambda({\bf r-R)})
\eeq
Substituting for ${\bf J}$ and $\rho$ using Eq.(\ref{2}) and expressing the coordinates ${\bf q}_{1}$ and ${\bf q}_{2}$
in terms of ${\bf R}$ and ${\bf q}$ using Eq.(\ref{3}) we have
\bea
L'=\frac{1}{2}M \dot{R}^2 + \frac{1}{2}\overline m \dot{q}^2 + \int d^3{\bf r} \left[
\frac{1}{2}\epsilon_0 \left\{{\bf E}^2({\bf r}) - c^2 {\bf B}^2({\bf r})\right\}
+{\bm {\cal P}}({\bf r}).{\bf E}({\bf r}) + \right. \nonumber \\
\left.{\bm {\cal M}}({\bf r}).{\bf B}({\bf r})
-\dot{\bf R}.\left\{{\bm {\cal P}}({\bf r})\times{\bf B}
({\bf r})\right\} \right],
\label{lag1}
\eea
where ${\overline m}=m_{1}m_{2}/M$ is the reduced mass. The vector fields ${\bm {\cal P}}({\bf r})$ and ${\bm {\cal M}}({\bf r})$ are, 
respectively, the electric polarisation and magnetisation associated with the charge and current sources of the atom, both
of which emerge here as closed integral forms representing the entire electric and magnetic multipole series to all orders
\beq
{\bm {\cal P}}({\bf r})=\sum_{i=1,2}\frac{e_i}{e}\;{\bm {\cal F}}({\bf q}_i,
{\bf r})
\eeq
\beq
{\bm {\cal M}}({\bf r})=\sum_{i=1,2}\frac{e_i}{e}\;{\bm \Theta}({\bf q}_i,
{\bf r})
\times({\bf {\dot q}}_i-{\bf {\dot R}})
=\left[\frac{m_2}{M}{\bm \Theta}({\bf q}_1,{\bf r})+
\frac{m_1}{M}{\bm \Theta}({\bf q}_2,{\bf r})\right]\times{\bf {\dot q}}
\eeq
Note that the magnetisation field vector of the two-particle system involves only orbital magnetic multipoles,
rather than the spin magnetic moment of the
particles,  but it is possible to generalise the theory to incorporate spin. 

The interpretation of each of the various terms in the gauge-transformed 
Lagrangian in Eq.(\ref {lag1}) is straightforward, except for
last term,
identified as the R\"{o}ntgen interaction term [7,11], which involves the
coupling between the centre of mass
motion,  the magnetic flux density and the electric polarisation field of
the system. 

The presence of the R\"{o}ntgen interaction term makes 
the Lagrangian unsymmetric between the
electric 
and magnetic properties,  specifically as far as the centre of mass 
motion is concerned. This lack of symmetry suggests that an interaction
term is missing which,  like the R\"{o}ntgen term, should be
first order in ${\bf {\dot R}}$ and which couples the magnetisation to the
electric field via the centre of mass motion.  The missing term is 
identified as the Aharonov-Casher term [12] and is obtainable by adding a
relativistic correction to the Lagrangian in Eq.(\ref{1}) so that
$\rho\rightarrow \rho'$ where
\beq
\rho'({\bf r})=\rho({\bf r})+\frac{1}{c^{2}}\dot{\bf R}.
\{\nabla \times{\bm {\cal M}}({\bf r})\}.
\label{4}
\eeq
Including this correction in the original Lagrangian, and following the
same
procedures as descibed above, leads to the appearance of a new term in the form of
the Aharonov-Casher term so that the new Lagrangian should be
\bea
L'=\frac{1}{2}M \dot{R}^2 + \frac{1}{2}{\overline m} \dot{q}^2 + \int d^3{\bf r} \left[
\frac{1}{2}\epsilon_0 \left({\bf E}^2({\bf r}) - c^2 {\bf B}^2({\bf r})\right)
+{\bm {\cal P}}({\bf r}).{\bf E}({\bf r})\right. \nonumber \\
\left.+{\bm {\cal M}}({\bf r}).{\bf B}({\bf r})
-\dot{\bf R}.\left\{{\bm {\cal P}}({\bf r})\times{\bf B}
({\bf r})-\frac{1}{c^{2}}{\bm {\cal M}}({\bf r})\times{\bf E}({\bf r})\right\} \right],
\label{lag'}
\eea
which is seen to be symmetric between electric and magnetic interactions.
Note, however, that the R{\"o}ntgen term was obtained without introduting
relativity while the Aharonov-Casher term required a relativistic correction.

\section{Single atom plus fields system}
\subsection{Canonical momenta}
The canonical variables in the transformed Lagrangian are the centre of mass coordinate ${\bf R}$
and internal coordinate ${\bf q}$ for the particle system and the vector and scalar potentials ${\bf {A}}$ and $\phi$
for the fields.  The corresponding canonical momenta are as follows.  For the
centre-of-mass motion  we have
\beq
{\bf P}=\frac{\partial L'}{\partial {\bf {\dot R}}}=M{\bf {\dot R}}-\int d^{3}{\bf r}\left\{{\bm {\cal P}}({\bf r})\times{\bf B}
({\bf r})-\frac{1}{c^{2}}{\bm {\cal M}}({\bf r})\times{\bf E}({\bf r})\right\}
\eeq
\label{L'}
while for the internal motion we have
\beq
{\bf p}=\frac{\partial L'}{\partial {\bf {\dot q}}}={\overline m}
{\bf {\dot q}}-\left[\frac{m_2}{M}{\bm \Theta}({\bf q}_1,{\bf r})+
\frac{m_1}{M}{\bm \Theta}({\bf q}_2,{\bf r})\right]\times
\left[{\bf B}({\bf r})-\frac{1}{c^2}{\bf {\dot R}}\times{\bf E}({\bf r})\right]
\eeq
For the fields we obtain the canonical momentum correponding to ${\bf A}$ as
\beq
{\bm {\Pi}}=\frac{\partial {\cal L}'}{\partial {\bf {\dot A}}}=
-\epsilon_{0}{\bf E}-{\bm {\cal P}}-\frac{1}{c^{2}}{\bf {\dot R}}\times{\bm {\cal M}}
\eeq
where ${\cal L}'$ is the Lagrangian density, identified in $L'$ as the
integrand in Eq. (\ref{lag'}).  The canonical momentum corresponding to $\phi$ 
vanishes.
\subsection{Constitutive relations}

The field equations follow from the Euler-Lagrange equations using $\phi$ and ${\bf A}$
as canonical variables.  They are identified as the familiar Maxwell equations 
\beq
{\bn {\nabla}}{\bf .D}=0;\;\;\;\;\;\;{\bn {\nabla}}\times{\bf H}=\frac{\partial {\bf D}}{\partial t}\label{max}
\eeq
where ${\bf D}$ is the electric displacement field 
and ${\bf H}$ is the magnetic field intensity which enter Maxwell's equations Eq.(\ref{max}) 
provided that the following constitutive
relations hold
\beq
{\bf H}({\bf r})=\frac{1}{\mu_{0}}{\bf B}({\bf r})-{\bm {\cal M}}+{\bf {\dot R}}\times{\bm {\cal P}}\label{cons2}
\eeq
\beq
{\bf D}({\bf r})=\epsilon_{0}{\bf E}({\bf r})+{\bm {\cal P}}+\frac{1}{c^{2}}{\bf {\dot R}}\times{\bm {\cal M}}\label{cons1}
\eeq
It is seen that these relations are themselves symmetric between the electric and magnetic contributions and
we note, in particular, that the last term in Eq.(\ref{cons1}) arises directly from the Aharanov-Casher Lagrangian
interaction term  The corresponding term in Eq.(\ref{cons2}) arises from the R\"{o}ntgen Lagrangian interaction term.

\subsection{Hamiltonian}

The Hamiltonian of the atomic system interacting with the electromagnetic fields now
follows from the Lagrangian by use of the canonical prescription
\beq
{\cal H}={\bf P.}{\bf {\dot R}}+{\bf p.}{\bf {\dot q}}+{\bn {\Pi .}}{\bf {\dot A}}-L'
\eeq
which yields,  after some manipulations,
\beq
{\cal H}=\frac{P^{2}}{2{\rm M}}+\frac{p^{2}}{2{\bar {\rm m}}}
+\frac{1}{2}\epsilon_0\int d^{3}{\bf r}\left\{\frac{1}{\epsilon_{0}^{2}}{\bn {\Pi}}^2({\bf r}) +
 c^2 {\bf B}^2({\bf r})\right\}+{\cal H}_{int}
\eeq
where ${\cal H}_{int}$ accounts for the coupling between the electromagnetic fields and the atomic system (ignoring diamagnetic terms)
\bea
{\cal H}_{int}&=&\int d^{3}{\bf r}\left(\frac{1}{2\epsilon_{0}}\{{\bn {\cal P}}({\bf r})\}^{2}
+\frac{1}{\epsilon_{0}}{\bm {\cal P}}({\bf r}).{\bm {\Pi}}({\bf r})
-{\bm {\cal M}}({\bf r}){\bf .B}({\bf r})\right)\nonumber\\
&+&\frac{1}{2M}\int d^{3}{\bf r}\;\left[{\bf P.}\left\{{\bm {\cal P}}({\bf r})\times{\bf B}({\bf r})
+\mu_0{\bm {\cal M}}({\bf r})\times{\bf \Pi}({\bf r})\right\}\right. \nonumber\\
&+&\left. \left\{{\bm {\cal P}}({\bf r})\times{\bf B}({\bf r})+\mu_0{\bm {\cal M}}({\bf r})\times{\bf \Pi}({\bf r})\right\}
{\bf .P}\right]\label{ham}
\eea
where the magnetisation field ${\bm {\cal M}}$ should now be symmetrised 
in terms of the internal coordinate and its canonical momentum.  Also in Eq.(\ref{ham}), pending 
further clarification, the term involving the square of the polarisation is deliberately included in the 
interaction Hamiltonian, although part of this term accounts for the Coulomb interaction between the particles.
  
 \section{Many-body formalism}
 
 \subsection{Quantum field theoretic Hamiltonian}

The single atom plus fields Hamiltonian now needs to be generalised to the many-body situation
involving an ensemble of atoms plus fields interacting quantum-mechanically. We begin by
introducing the boson field opertor ${\hat {\Psi}}$ describing the internal as well as the
centre-of mass motion of the ensemble, written as a sum over product eigenstates as follows
\beq
{\hat {\Psi}}({\bf q,R})={\hat {\psi}}({\bf R}){\hat {\chi}}({\bf q})
=\sum_{n_{1},n_{2}}\psi_{n_{1}}({\bf R})\chi_{n_{2}}({\bf q}){\hat a}_{n_{1},n_{2}}
\eeq
where $\psi_{n_{1}}({\bf R})$ are state functions associated with the centre of
mass and $\chi_{n_{2}}({\bf q})$ with the internal atomic motion.  The labels $n_{1}$ and $n_{2}$
incorporate all quantum numbers specifying these states.  The operators 
${\hat a}_{n_{1},n_{2}}$ and ${\hat a}^{\dagger}_{n_{1},n_{2}}$ are annihilation and creation opertors satisfying 
bosonic commutation relations.
Completeness demands that we must have
\beq
\sum_{n_{1}}\psi^{*}_{n_{1}}({\bf R})\psi_{n_{1}}({\bf R'})=\delta({\bf R-R'});\;\;\;\;\;\;\;\sum_{n_{2}}\chi^{*}_{n_{2}}({\bf q})\chi_{n_{2}}({\bf q'})=\delta({\bf q-q'})
\eeq
In the next step the many-body Hamiltonian operator is written as follows
\bea
{\hat H}&=&\int\;d^{3}{\bf q}d^{3}{\bf R}{\hat {\Psi}}^{\dagger}({\bf q,R})\left(\frac{P^{2}}{2M}+\frac{p^{2}}{2{\bar m}}\right){\hat {\Psi}}({\bf q,R})
+\frac{1}{2}\epsilon_0\int d^{3}{\bf r}\left\{\frac{1}{\epsilon_{0}^{2}}{\bm {\Pi}}^2({\bf r})+c^2 {\bf B}^2({\bf r})\right\}\nonumber\\
&+&\int\;d^{3}{\bf q}d^{3}{\bf R}{\hat {\Psi}}^{\dagger}({\bf q,R})V_{trap}{\hat {\Psi}}({\bf q,R})\nonumber\\
&+&\frac{1}{2}\int\;d^{3}{\bf q}d^{3}{\bf R}\int\;d^{3}{\bf q'}d^{3}{\bf R'}{\hat {\Psi}}^{\dagger}({\bf q,R}){\hat {\Psi}}^{\dagger}({\bf q',R'})U({\bf R-R'})
{\hat {\Psi}}({\bf q,R}){\hat {\Psi}}({\bf q',R'})\nonumber\\
&+&{\cal H}^{Q}_{int}\label{m-b}
\eea 
where ${\cal H}^{Q}_{int}$  is given by
\bea
{\cal H}^{Q}_{int}&=&\int d^{3}{\bf r}\left[\frac{1}{2\epsilon_{0}}\{{\hat {\bn {\cal P}}}({\bf r})\}^{2}
+\frac{1}{\epsilon_{0}}{\hat {\bm {\cal P}}}({\bf r}).{\bm {\Pi}}({\bf r})
-{\hat {\bm {\cal M}}}({\bf r}).{\bf B}({\bf r})\right]\nonumber\\
&+&\frac{1}{2M}\int d^{3}{\bf q}d^{3}{\bf R}\int d^{3}{\bf r}\;{\hat {\Psi}}^{\dagger}({\bf q,R})\left[{\bf P}.
\left\{{\hat {\bm {\cal P}}}({\bf r})\times{\bf B}({\bf r})
+\mu_0{\hat {\bm {\cal M}}}({\bf r})\times{\bf \Pi}({\bf r})\right\}\right.\nonumber\\
&+&\left.\left\{{\hat {\bm {\cal P}}}({\bf r})\times{\bf B}({\bf r})+\mu_0{\hat {\bm {\cal M}}}({\bf r})\times{\bf \Pi}({\bf r})\right\}
{\bf .P}\right]{\hat {\Psi}}({\bf q,R})
\eea
with the polarisation and magnetisation
converted to field theoretic operators as follows:
\beq
{\hat {\bn {\cal P}}}=\sum_{i=1,2}e_{i}\int\;d^{3}{\bf q}d^{3}{\bf R}{\hat {\Psi}}^{\dagger}({\bf q,R})\int_{0}^{1}d\lambda({\bf q}_{i}-{\bf R})
\delta({\bf r-R}-\lambda({\bf q}_{i}-{\bf R}){\hat {\Psi}}({\bf q,R}) 
\eeq
\beq
{\hat {\bn {\cal M}}}=\sum_{i=1,2}\frac{1}{2m_i}\int\;d^{3}{\bf q}d^{3}
{\bf R}{\hat {\Psi}}^{\dagger}({\bf q,R})
\left[{\bm \Theta}({\bf q}_i,{\bf r}) \times{\bf p}-
{\bf p}\times{\bm \Theta}({\bf q}_i,{\bf r})\right]
{\hat {\Psi}}({\bf q,R})
\eeq
Note that although the polarisation and magnetisation operators are expressed above in terms of ${\bf q}_{1}$
and ${\bf q}_{2}$,  they can readily be written in terms of the relative and centre-of mass coordinates using Eq.(\ref{3}).

We have introduced two new terms in the many-body Hamitonian Eq.(\ref{m-b}).  The term involving $V_{trap}$ accounts for
the trapping potential used to confine the atomic ensemble,  and the term involving $U({\bf R-R'})$ is identified as
the hard sphere collision term.  This is usually taken to be of the form 
\beq
U({\bf R-R'})=\frac{2\pi\hbar^{2}a}{M}\delta({\bf R-R'})\equiv U_{0}\delta({\bf R-R'})\label{30}
\eeq
where $a$ is the scattering length.

\subsection{Generalised Gross-Pitaevskii equation}

Having arrived at the appropriate many-body quantum field theoretical Hamiltonian we can now derive the Schr\"{o}dinger equation 
satisfied by the atomic field ${\hat {\Psi}}({\bf q,R})$.  This formally follows from the Heisenberg equation 
\beq
i\hbar{\dot {\hat {\Psi}}}=\left[{\hat H}\;,\;{\hat {\Psi}}\right]=E{\hat \Psi}\label{heis}
\eeq
It is instructive to check first what the outcome would be in the absence of all electromagnetic terms plus
terms involving the polarisation.  In this case Eq.(\ref{heis}) yields
\beq
\left[\frac{P^{2}}{2M}+\frac{p^{2}}{2{\bar m}}+V_{trap}+U_{0}{\hat {\Psi}}^{\dagger}({\bf q,R}){\hat {\Psi}}({\bf q,R})\right]{\hat {\Psi}}({\bf q,R})
=E{\hat {\Psi}}({\bf q,R})\label{gp1}
\eeq
Apart from the implicit dependence on the internal quantum states $\chi({\bf q})$ and the explicit
appearance of the kinetic energy term $p^{2}/2{\bar m}$, Eq.(\ref{gp1}) is reminiscent of the Gross-Pitaevskii
equation prior to assuming that all ensemble constituents occupy the ground quantum state.

On reintroducing the electromagnetic interaction terms the full Schr\"{o}dinger equation obtained turns out to be a modified
 Goss-Pitaevskii equation.  The modifications 
involve the presence of the internal states, including their coupling to the gross motion and the
 coupling of the two motions to the electromagnetic fields.  In order to derive this general equation
 we need to discuss the origins of various terms which will appear in it.  First there must be a term
 accounting for the binding of the particles within a single atom,  i.e. intra-atom Coulomb interactions,
 and there must also be terms accounting for the interaction between the atoms,  i.e. the inter-atom
 interactions.  Both Coulmb interactions arise from the field theoretic  term
 \beq
 {\hat H}_{c}=\frac{1}{2\epsilon_{0}}\int d^{3}{\bf r}\left([{\hat {\bn {\cal P}}_{intra}}^{\parallel}]^{2}
 +[{\hat {\bn {\cal P}}_{inter}}]^{2}\right)
 \eeq
We have dropped the additional term involving the transverse part of the intra-atom polarisation,  as this can be shown
to lead to infinite self energies.  The full inter-atom polarisation term gives the electromagnetic interactions between the
atoms.

Assuming that inter-atomic separations $|{\bf R-R'}|$ are typically much larger than a dipole length $|{\bf q}|$,
which amounts to the dipole approximation, we can write
\beq
{\bm {\cal P}}=N{\bf d}|{\psi}({\bf R})|^{2};\;\;\;\;\;\;{\bm {\cal M}}=N{\bn {\mu}}|{\psi}({\bf R})|^{2}
\eeq
where $N$ is the number of atoms in the ensemble,  ${\bf d}$ is the electric dipole moment and ${\bn {\mu}}$
is the magnetic dipole moment.  The modified Gross-Pitaevskii equation in the dipole
approximation turns out to be in the form
\beq
\left[\frac{P^{2}}{2M}+V_{trap}+\frac{p^{2}}{2{\bar m}}-\frac{e^{2}}{4\pi\epsilon_{0}q}
+U_{0}{\hat {\Psi}}^{\dagger}({\bf q,R}){\hat {\Psi}}({\bf q,R})+V_{dd}+V_{em}\right]{\hat {\Psi}}({\bf q,R})
=E{\hat {\Psi}}({\bf q,R})
\eeq
where $V_{dd}$ is an inter-atom interaction term
\beq
V_{dd}=\frac{e}{\epsilon_{0}}\int d^{3}{\bf q'}({\bf d.q'})\left\{2{\hat {\Psi}}^{\dagger}({\bf q',R}){\hat {\Psi}}({\bf q',R})-
{\hat {\Psi}}^{\dagger}({\bf q',R}){\hat {\Psi}}({\bf q',R})\right\}
\eeq
while $V_{em}$ includes the coupling of the electric and magnetic dipole systems, pertaining to the internal dynamics, 
with the centre of mass and the electromagnetic fields
\bea
V_{em}&=&-{\bf d.E}^{\bot}({\bf R})-{\bn {\mu}}{\bf .B}({\bf R})\nonumber\\
&+&\frac{1}{2M}\left[{\bf P.}\{{\bf d}\times {\bf B}({\bf R})
-\frac{1}{c^{2}}{\bm {\mu}}\times {\bf E}({\bf R})\}+\{{\bf d}\times {\bf B}({\bf R})
-\frac{1}{c^{2}}{\bm {\mu}}\times {\bf E}({\bf R})\}{\bf .P}\right]
\eea
Next we assume that the number of atoms in the BEC is very large and that they all occupy the ground state.
Under such circumstances only the ground state operators are involved and these become c-numbers such that
\beq
{\hat a}_{0,0}\approx\sqrt{N};\;\;\;\;{\hat a}^{\dagger}_{0,0}\approx\sqrt{N}
\eeq
All terms involving other operators ${\hat a}_{n_{1},n_{2}}$ and  ${\hat a}^{\dagger}_{n_{1},n_{2}}$ where $n_{1}\neq 0$
and $n_{2}\neq 0$ give vanishing results.  Although this greatly simplifies the problem,  the internal and the centre-of-mass 
motions are still coupled.  To achieve a decoupling of the two motions we need to perform enemble averaging.  This allows us to write
\beq
 V_{dd}=\frac{e}{\epsilon_{0}}N|\psi_{0}({\bf R})|^{2}\int d^{3}{\bf q'}{\bf d.q'}\left|\chi_{0}({\bf q'})\right|^{2} 
\eeq
The ensemble averaging also results in the substitution
\beq
E=E_{\mu}+{\cal E}_{0}
\eeq
where $E_{\mu}$ is the chemical potential  and ${\cal E}_{0}$ is the lowest internal energy, i.e.
 the ground state energy eigenvalue of Schr\"{o}dinger's equation 
for the internal motion
\beq
\left\{\frac{p^{2}}{2{\bar m}}-\frac{e^{2}}{4\pi\epsilon_{0}q}\right\}\chi_{0}({\bf q})={\cal E}_{0}\chi_{0}({\bf q})
\eeq
with $\chi_{0}({\bf q})$ the corresponding hydrogenic ground state eigenfunction.  The final step is to assume that in a condensate there 
should be a maximum correlation between the
atoms, in which case we can  identify $V_{dd}$ as an effective dipole-dipole interaction in the form
\beq
V_{dd}=\frac{d^{2}}{\epsilon_{0}}N|\psi_{0}({\bf R})|^{2}
\eeq
The modified Gross-Pitaeveskii equation now becomes
\beq
\left[\frac{P^{2}}{2M}+V_{trap}+N\left|\psi_{0}({\bf R})\right|^{2}\left(\frac{d^{2}}{\epsilon_{0}}+U_{0}\right)+V_{em}\right]{\hat {\Psi}}({\bf q,R})
=E_{\mu}{\hat {\Psi}}({\bf q,R})\label{grosspit}
\eeq 
Like the Gross-Pitaevskii equation in the absence of electromagnetic effects,  this is a non-linear Schr\"{o}dinger equation.
The additional terms are $V_{dd}$, the dipole-dipole interaction, and $V_{em}$, the electromagnetic interaction.  All the new terms have
been arrived at systematically within our theory,  including the dipole-dipole term which in previous considerations has been 
added phenomenologically.  To our knowledge the new interaction terms included in $V_{em}$ have not been derived rigorously before.
To see the effects of the electromagnetic fields on the properties of a BEC we 
now consider the special case of a BEC in a vortex state

\section{Rotating BECs}

In view of the typical experimental arrangements for generating BECs we may assume that the BEC is confined within
a cylinderical region $|z|<z_{0}$ with cylinder radius $R_{0}$ and height  $2z_{0}$.  The length $z_{0}$ can be larger or 
smaller than $R_{0}$ depending on
the type of trap, but below we shall consider the cases where
$z_{0}\approx R_{0}$ and $z_{0}$ infinite.
We perform a Madelung transformation [6] and,
assuming that the wavefunction is independent of $z$, write
\beq
\psi({\bf R})=\left(\frac{1}{4\pi z_0 R_0^2}\right)^{1/2}|\psi(r)|e^{in\phi}
\eeq
where $(r,\phi,z)$ are the cylindrical coordinates of ${\bf R}$.
On considering solutions of Eq. (\ref {grosspit}) we may drop the electromagnetic
field interaction term, which introduces small second order corrections [7],
and the dipole-dipole interaction term which is also negligible compared to 
the s-wave scattering term.
We thus have for the velocity profile of the BEC in an order n vortex state
\beq
\dot{\bf R}=\left(\frac{n\hbar }{Mr}\right)\hat{\bm \phi}
\eeq 
and also the radial wavefunction satisfies the equation [7]
\beq
\left[\frac{d^2}{d\xi^2}+\frac{1}{\xi}\frac{d}{d\xi}-\frac{n^2}{\xi^2}+
\frac{2ME}{\hbar^2}-4n_{1d}a\psi(\xi)|^2\right]\psi(\xi)=0,
\label{8}
\eeq
with $a$ the scattering length first encountered in Eq.(\ref{30}); $\xi=r/R_0$, $n_{1d}=N/2z_0$ is the one-dimensional atomic density
parameter and we set  $\psi(\xi=1)=0$.
We note that the solution to this equation
for $n_{1d}a=0$ would be the Bessel funtion of the first kind, $J_n(x)$, with
a root at $\xi=1$.
Figure 1(a) displays the modulas squared of the
numerical solution of Eq.(\ref{8}) for various values of  $n_{1d}a$.

\subsection{Electric BEC}

In an electric BEC the constituent atoms are characterised by an electric dipole moment.
We assume that the moment vectors all point in the $z$-direction. We concentrate on the
static case and , setting ${\bn {\cal M}}=0$, we take the divergence of both sides in Eq. (\ref{cons2}).
The divergence of the magnetic flux density ${\bf B}$ is automatically zero.
Hence,
\bea
\nabla.{\bf H}=\nabla.(\dot{\bf R}\times{\bm {\cal P}})&=&
{\bm {\cal P}}.(\nabla\times\dot{\bf R})-
\dot{\bf R}.(\nabla\times{\bm {\cal P}}) \nonumber \\
&=&\frac{\hbar n}{Mr}\left[\delta(r)+\frac{d}{dr}\right]
\frac{n_{1d}d}{2\pi R_0^2}|\psi(r)|^2
=\frac{\rho_m(r)}{\mu_0}.
\label{47}
\eea
where we have expressed the right hand side  in the first equality of Eq. (\ref {47}) as a
magnetic monopole charge density divided by $\mu_{0}$. The magnetic field intensity can
be written as the gradient of a scalar field: ${\bf H}=-\nabla\Phi$, where
$\Phi$ is the magnetic vector potential.
This is similar to the result obtained in Ref.[7]. However, in [7] it was
argued that, although the wavefunction vanishes on the cylinder axis, some
atoms will leak into the core of the vortex and this gives rise to the monopole
charge. The  variations of the wavefunction across the cylinder were
ignored. In solving Eq. (\ref{47}) we note that the term involving the delta
function in the first equality cancels with an identical term arising from 
integrating the derivative which is proportional to $|\psi(0)|^2$.
This feature of our theory guarantees  `charge neutrality', in contrast with the case 
in Ref.[7].

Figure 1(b) displays the variation of the derivative of the modulus squared of
the wavefunction. This quantity is proportional to the two-dimensional monopole charge 
density in a symmetry plane containing the cylinder axis and can be written in terms of $\xi=r/R_{0}$
as follows
\beq
2\pi\xi\rho_m(\xi)=\frac{\hbar n n_{1d}\mu_0 d}{MR_0^4}\;\left(\frac{d|\psi(\xi)|^2}{d\xi}\right)
\eeq
Clearly as the scattering length becomes longer or the number of
atoms increases the
wavefunction broadens and the associated charges accumulate near the
axis and edge of the cylinder. However,  the integral of the density
over the cylinder volume reduces to zero, as demanded by charge neutrality.
This magnetic monopole charge distribution can be thought of as akin to
that found in a cylindrical capacitor with one set of charges on an inner thin
cylinder and another set of charges of opposite sign on an outer
cylinder. A useful measure of the effect is the cumulative positive charge, $Q_m$,
in the inner region of the cylinder, which can be estimated
by integrating the charge density up to $r=R_0/2\; (\xi=1/2)$.  We find 
\beq
Q_m=\frac{\hbar n N \mu_0 d} { MR_0^2}|\psi(\xi=1/2)|^2\label{49}
\eeq
For an infinite cylinder, Eq. (\ref{47}) can be solved using
Gauss's law to obtain
\beq
{\bf H}(\xi)=\hat{\bf r}\frac{\hbar n n_{1d} d}{2\pi M R_0^3
\xi}\int_0^\xi
\left[\delta(\xi')+\frac{d}{d\xi'}\right]|\psi(\xi')|^2 d\xi'=
\hat{\bf r}\frac{\hbar n n_{1d} d}{2\pi M R_0^3 \xi}|\psi(\xi)|^2.
\label{50}
\eeq
where ${\hat {\bf r}}$ is a cylindrical radial unit vector.
We see that ${\bf H}$ is proportional to $|\psi(\xi)|^2$ which vanishes outside the cylinder. There is a
magnetic field only inside the cylinder, another signature of the conservation
of `magnetic monopole charge'.

If we wish to detect the magnetetic effects associated with the vortex, it may be 
necessary to probe the region outside
the BEC which can have a magnetic field for a cylinder of a finite height
$2z_{0}$. The solution for the magnetic potential in the finite cylinder case can be derived using Green
 functions [18]. We find
\beq
\Phi(\xi,z)=\Phi_0^m \int_0^1 d\xi' \int_0^{2\pi} d\phi'
\frac{\xi'-\xi \cos(\phi')}{h^2}\left[\frac{1-z}{(h^2f^2+(1-z)^2)^{1/2}}
+\frac{1+z}{(h^2f^2+(1+z)^2)^{1/2}}\right],
\label{51}
\eeq
where $\Phi_{0}^{m}$ is a scaling magnetic potential controlling the order of magnitude of the 
effect
\beq
\Phi_0^m=\frac{\hbar n n_{1d} d}{8\pi^2 M R_0^2}\label{52}
\eeq
the variable $z$ is in units of $z_0$ and we have used the notation
\beq
 h^2=\xi^2+\xi'^2-2\xi\xi'\cos(\phi'),\;\;\;\;
f=\frac{R_0}{z_0}
\eeq
Figure 2 displays a contour plot of the magnetic potential on a symmetry plane
through the cylinder axis and above the symmetry plane of $z=0$.
As expected, because of the finite height of the
cylinder, the magnetic potential is not confined to the inside  but
leaks to the  outside of the cylinder.
Within the cylinder the magnetic potential exhibits its largest variations where
the wavefunction changes most rapidly. 

It is useful to obtain
order of magnitude estimates of the effects just described for the cases of a typical
atomic gas BEC and for superfluid helium.
In the case of an atomic gas BEC, we choose typical parameters
appropriate for $^{87}$Rb [5] in the $n=1$
vortex state, assuming that the BEC is confined in a cylinder of dimensions $R_0=z_0=2\mu$m.
Each atom is assumed to be characterised by a transition electric dipole moment
$d=e a_B$, where $a_B=0.53$\AA$\;$ is the Bohr radius, and the
s-wave scattering length is taken to be $a=59$\AA. Setting $n_{1d}a=100$
gives a linear
density of $n_{1d}=1.7\times10^{10}$m$^{-1}$ and a total number of atoms
$N=7\times10^4$. Thus we have for $\Phi_{0}^{m}$
\beq\Phi_0^m=3.3\times10^{-19}\;{\rm A}\label{54}
\eeq
and for the positive magnetic monopole charge residing in the inner region of the cylinder, Eq. (\ref{49})
\beq 
Q_m=1.3\times10^{-28} |\psi(\xi=1/2)|^2\;{\rm {Vs}}\label{55}
\eeq
 which is equivalent to a magnetic field of order $10^{-19}$T. 
 
In the case of superfluid helium, which is taken to be  characterised by macroscopic electric susceptibility 
the polarisation field vector can be written as
\beq
{\bm {\cal P}}(\xi)=\chi{\bf E}|\psi(\xi)|^2
\eeq
 where ${\bf E}$ is an applied electric field and $\chi=0.052$ is the value of the susceptability. We have, in effect
replaced the volume density times the dipole moment with a susceptability times an
externally applied electric field. Thus, for every $1$Vm$^{-1}$ of applied field the magnetic monopole charge 
in the inner region of the cylinder is
\beq
Q_m=1.2\times10^{-20} |\psi(\xi=1/2)|^2\;{\rm {Vs}}\label{56}
\eeq
 and a magnetic field of order $10^{-14}$T (equivalent to the values in [7]).
This larger magnitude stems from the smaller helium mass and larger density (approximately
$10^{21}$m$^{-3}$ in the case of the atomic BEC above and $10^{28}$m$^{-3}$ for superfluid
helium). The potential estimated in Eq.(\ref{54}) for the Rb atomic BEC is clearly too small
to be measured on the basis of current experimental capabilities.
However, one expects that denser atomic gas BECs will
become available in the future. The corresponding estimate found above in the case of superfluid
helium indicates that the effects are, indeed amenable to experimental detection, as deduced in [7].

\subsection{Magnetic BEC}

Next we consider to the case of a BEC in which the constituent
atoms are characterised by magnetic dipoles aligned along the $z$-direction.
Setting ${\bn {\cal P}}=0$,  we take the divergence of both sides in Eq. (\ref {cons1}),
and following the procedure in the electric BEC  case, we
obtain
\bea
\epsilon_0\nabla.{\bf E}=-\frac{1}{c^2}
\nabla.(\dot{\bf R}\times{\bm {\cal M}})&=&
-\frac{1}{c^2}\left[{\bm {\cal M}}.(\nabla\times\dot{\bf R})-
\dot{\bf R}.(\nabla\times{\bm {\cal M}})\right] \nonumber \\
&=&-\frac{\hbar n}{c^2Mr}\left[\delta(r)+\frac{d}{dr}\right]
\frac{n_{1d}\mu}{2\pi R_0^2}|\psi(r)|^2
=\rho_e(r).
\label{58}
\eea
It can be seen that, for the fields and potentials,
Eq.(\ref{58}) transforms to Eq.(\ref{47})
 when $d$ is replaced by $-\mu_0\mu$. Therefore we can deduce
the electric field associted with the vortex state of an infinitely 
long cylinderical BEC directly by simple substitution.  The electrostatic potential
is related to the electric field by ,
${\bf E}=-\nabla \Phi$ and is given directly by by Eq.(\ref {51}) and (\ref{52}), with $\Phi_0^m$
replaced by
\beq
\Phi_0^e=-\frac{\hbar n n_{1d}\mu_0 \mu}{ 8\pi^2 M R_0^2}
\eeq
Figure 1(b) now serves to display a function proportional to the 
areal electric charge density in a symmetry plane such that
\beq
2\pi\xi  \rho_e(\xi)=-\frac{\hbar n n_{1d}\mu}{M c^2 R_0^4}\;\left(\frac{d|\psi(\xi)|^2}{d\xi}\right)
\eeq
Figure 2 should now be taken to present a function proportional to the electrostatic potential. The 
corresponding cumulative negative electric charge residing in the inner region of the cylinder 
obtained by integrating $\rho_{e}$ over the cylinder volume is approximately
\beq
Q_e=-\frac{\hbar n N \mu}{ Mc^2R_0^2}\;|\psi(\xi=1/2)|^2
\eeq

As to the order of magnitude, the best candidate for observing this effect 
is spin-polorised hydrogen [10] which has been
produced as a BEC. The hydrogen is doubly spin-polarised, thus
$\mu=2\mu_b=\hbar e/m_e=1.9\times10^{-23}$Am$^2$, a typical density
is $5\times10^{21}$m$^{-3}$ and the s-wave scattering length
is $a=0.72$\AA$\;$. Again assuming that $n_{1d}a=100$ and
$z_0=5$mm  we obtain 
\beq
\Phi_0^e=-2.7\times10^{-16}\;{\rm V}\;\;{\rm {and}}\;\;\;
Q_e=-1.9\times10^{-27}|\psi(\xi=1/2)|^2{\rm  {C}}
\eeq

Clearly these magnitudes are too small for the effect to be experimentally detectable at present.
However, the effect should become amenable to measurement whenever denser
spin-polarised hydrogen BECs can be produced (with densities of the order of 
those of superfluid helium).
\newpage

\section{Comments and Conclusions}

We have presented a rigorous treatment of the electromagnetic properties of 
BECs excuting rotational motion when in an order n vortex state.  The theory
is based on a Lagrangian which we have constructed in such a manner as to emphasise 
the symmetry of the problem with regards to the centre of mass motion.
The Lagrangian led us to constitutive relations which link the fields
to the polarisation sources, including the convective contributions arising from the
centre of mass motion.  We have also shown how direct generalisation to the many-body
situation can be done,  leading to a modified Gross-Pitaevskii equation which
incorporates familiar interactions plus the coupling to the electromagnetic fields.
Applications of the theory to the case of a rotating BEC is straightforward 
and we have shown how this leads to the electric and magnetic properties of a
rotating BEC,  depending on whether the constituent atoms are electric dipole active 
or magnetic dipole active. 

In particular, we have seen how a dipole-active BEC generates electric
and
magnetic monopole
charge distributions and their associated  electric and magnetic fields
when the
condensate is in a vortex state. We have also shown how, even if no
condensate
atoms occupy the vortex core, monopole charge distributions, both
electric
and magnetic, arise which are
globally neutral. The problem becomes similar to that of  finding the
potential in
a cylindrical capacitor. In the case of an atomic gas BEC of finite size
of either kind, the vortex state generates fields outside it
which are small when calculated on the basis of current estimates.
Their experimental detection will have to await the advent of denser
atomic gas
BECs than are currently available. The predicted magnetic fields
are much larger for a denser and more electrically
polarisable system such as a superfluid. In the case of superfluid
helium, we have shown that the magnetic distributions are sufficiently
large
to be amenable to experimental detection. It is envisaged that vortices generated in future
denser BECs of spin-polarised hydrogen would give rise to electric
distributions
which could be measurable.

\section*{Acknowledgments}

We thank Edward Hinds, University of Sussex, Gerald Hechenblaikner,
University of Oxford and Lincoln Carr, University of Washington
for useful discussions.
C.R.B. would like to thank the EPSRC for financial support and L.G.B.
would
like to thank the University of York for financial support. This work has
been
carried out under the EPSRC Grant No. GR/M16313.

\section*{References}

\begin{enumerate}
\item M. H. Anderson, J. R. Ensher, M. R. Matthews, C. E. Wieman and
E. A. Cornell, Science {\bf 269} 198 (1995); C. C. Bradley, C. A Sackett,
J. J. Tollett and R. G. Hulet, Phys. Rev. Lett. {\bf 75} 1687 (1995);
K. B. Davies, M.-O. Mewes, M. R. Andrews, N. J. van Druten, D. S. Durfee,
D. M. Kurn and W. Ketterle, Phys. Rev. Lett. {\bf 75} 3969 (1995).
\item F. Dalfovo, S. Giorgini, L. P. Pitaevskii and S. Stringari, Rev.
Mod. Phys. {\bf 71} 463 (1999).
\item D. M. Stamper-Kurn, M. R. Andrews, A. P. Chikkatur, S. Inouye,
H.-J.
Miesnerm, J. Stenger and W. Ketterle, Phys. Rev. Lett. {\bf 80} 2027
(1998);
E. M. Wright, J. Arlt and K. Dholakia, Phys. Rev. A {\bf 63} 013608
(2000).
\item M. D. Barett, J. A. Sauer and M. S. Chapman,  Phys. Rev. Lett. {\bf 87}, 010404 (2001)
\item A. L. Fetter, "Theory of a dilute low-temperature Trapped
Bose-Einstein
condensate" LANL Archive cond-mat-9811366 (1998);
F. Dalfovo and S. Stringari, Phys. Rev. A {\bf 53} 2477 (1996); and
references
therein.
\item R. J. Donnelly, "Quantized Vortices in Helium II"
(Cambridge University Press 1991).
\item U. Leonardt and P. Piwnicki, Phys. Rev. Lett. {\bf 82} 2426
(1999).
\item K. W. Madison, F. Chevy, W. Wohlleben and J. Dalibard, Phys. Rev.
Lett.
{\bf 84} 806 (2000);
K. G. Petrosyan and L. You, Phys. Rev. A {\bf 59} (1999);
E. V. Goldstein, E. M. Wright and P. Meystre, Phys. Rev. A {\bf 58}
576 (1998).
\item S I Shevshenko, J. Low Temperature Phys. {\bf 121}, 429 (2000)
\item T. J. Greytak, D. Kleppner, D. G. Fried, T. C. Killian, L.
Willmann,
D. Landhuis and S. C. Moss, Physica B {\bf 280} 20 (2000); T. Willmann,
Appl. Phys. B: Lasers and Optics {\bf 69} 357 (1999); T. J. Greytak in "Bose-Einstein Condensation", Ed. A. Griffin, D.
W.
Snoke and S. Stringari (CUP, Cambridge 1996); I. F. Silvera, {\it ibid}.
\item W. C. R\"{o}ntgen, Ann. Phys. Chem. {\bf 35} 264 (1888);
M. Babiker, E. A. Power and T. Thirunamachandran, Proc. Roy. Soc. A
{\bf 332} 187 (1973);
M. Wilkens, Phys. Rev. Lett. {\bf 72} 5 (1994).
\item Y. Aharonov and A. Casher, Phys. Rev. Lett. {\bf 53} 319 (1984).
\item K. Sangster, E. A. Hinds, S. M. Barnett, E. Riis and A. G.
Sinclair,
Phys. Rev. A {\bf 51} 1776 (1995).
\item V. E. Lembessis, M. Babiker, C. Baxter and R. Loudon, Phys. Rev.
A
{\bf 48} 1594 (1993).
\item C. R. Bennett, L. G. Boussiakou and M. Babiker, Phys. Rev. A, in press (2001).
\item M. Babiker and R. Loudon, Proc. R. Soc. Lon. A {\bf 385} 439
(1983).
\item C. Cohen-Tannoudji, J. Dupont-Roc and G. Grynberg, "Photons and
Atoms,
Introduction to Quantum Electrodynamics" (Wiley, New York 1989).
\item G. Barton, "Elements of Green's Functions and Propagation"
(Clarendon Press, Oxford 1989); P. M. Morse and H. Feshbach,
"Methods of Theoretical Physics" (McGraw-Hill, New York 1953).
\end{enumerate}

\newpage
\section*{Figure Captions}

\subsection*{\bf Figure 1}

Variations of (a) the modulus squared of the normalised BEC
wavefunction in the $n=1$ vortex state and (b) its derivative with
$n_{1d}a=100$ (solid curve), $n_{1d}a=10$ (dashed curve) and
$n_{1d}a=0.1$ (dotted curve).

\subsection*{\bf Figure 2}

A coutour plot within a symmetry plane of the cylinder showing the
variations of the electric (magnetic) potential of a magnetic (electric) dipole active BEC in the
$n=1$ vortex state which occupies the shaded region. The potential exhibits angular symmetry around $R=0$, is symmetric about the plane $z=0$
and is in units of $\Phi_0^e$ for the electric potential and $\Phi_0^m$ for
the magnetic potential (see text). The condensate is such that $z_0/R_0=1$ and $n_{1d}a=100$.

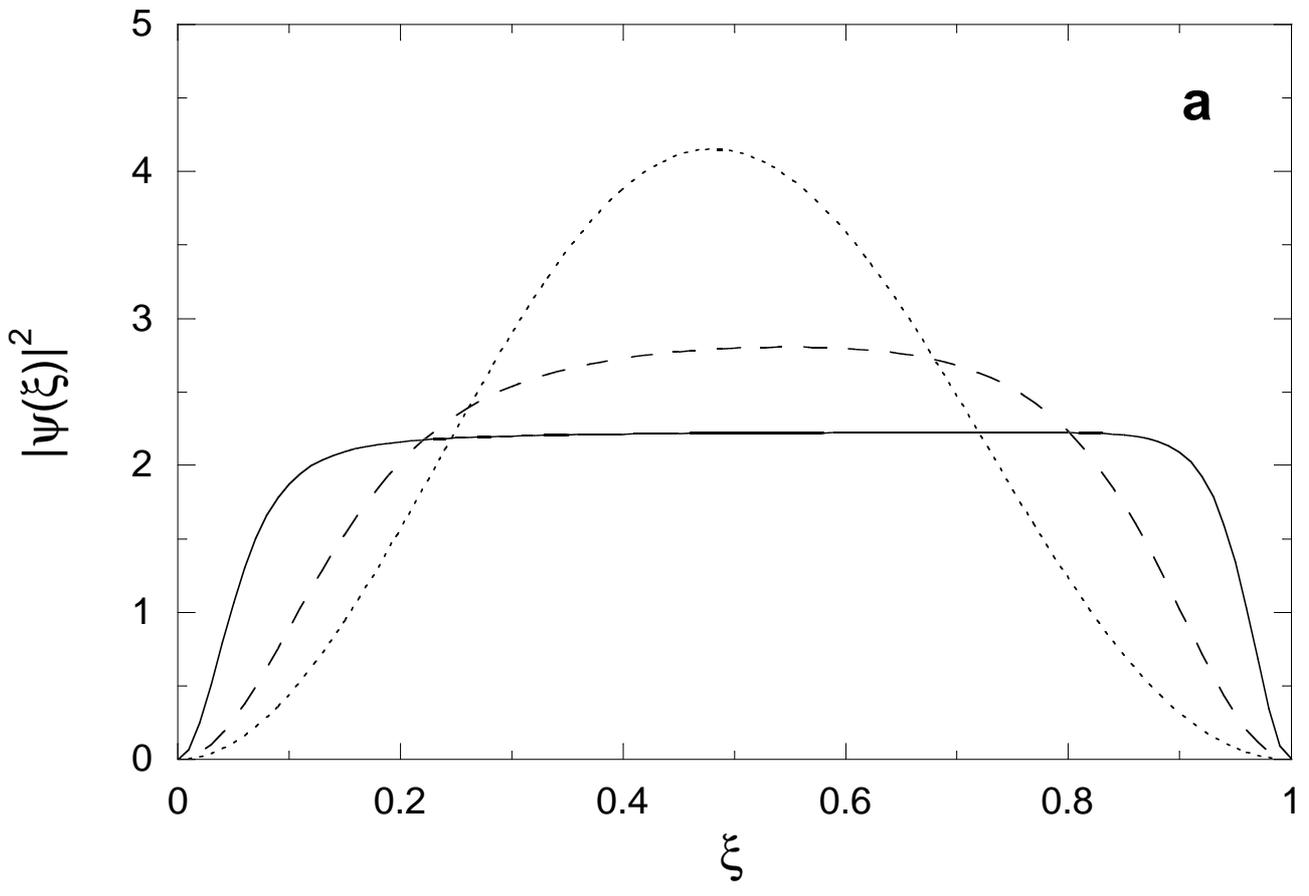
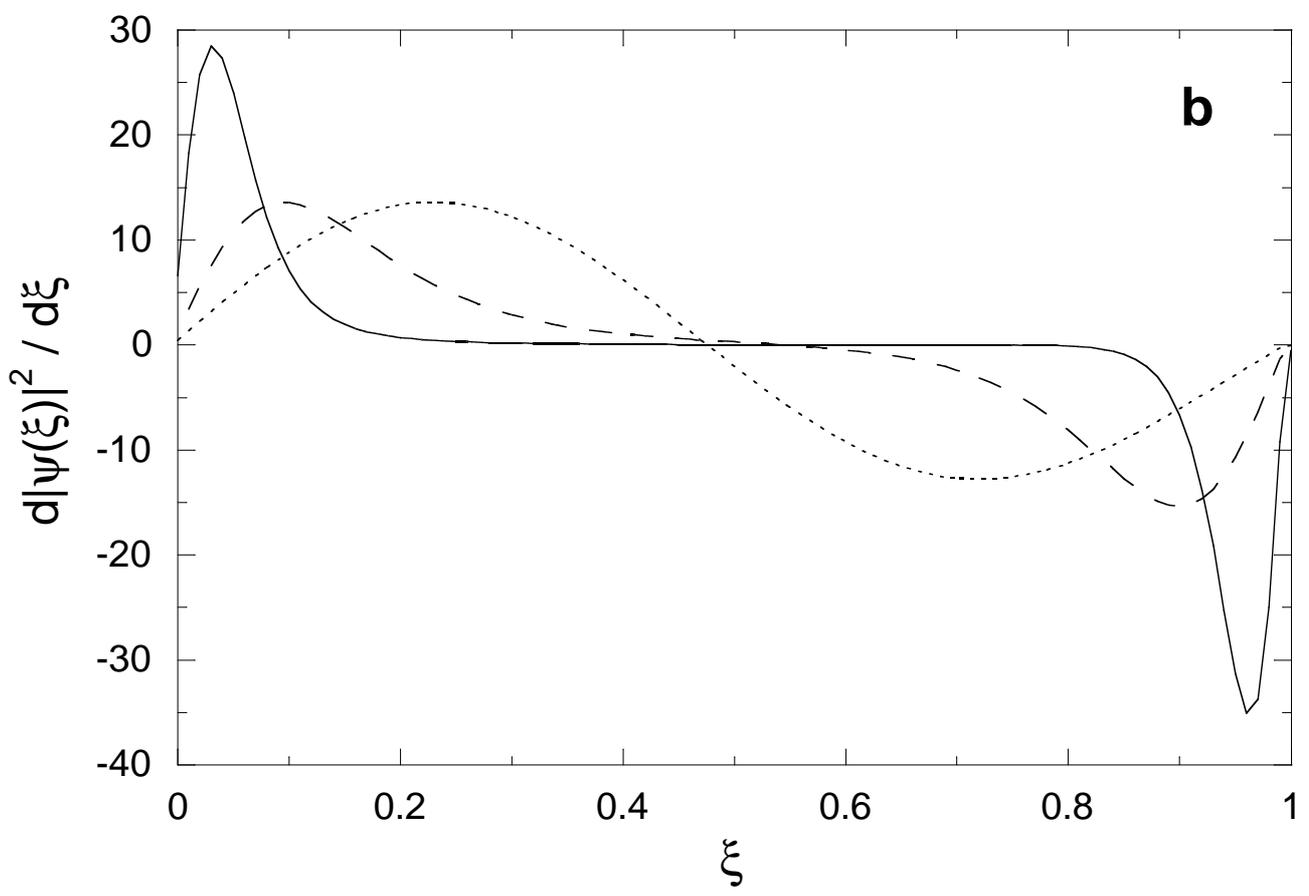
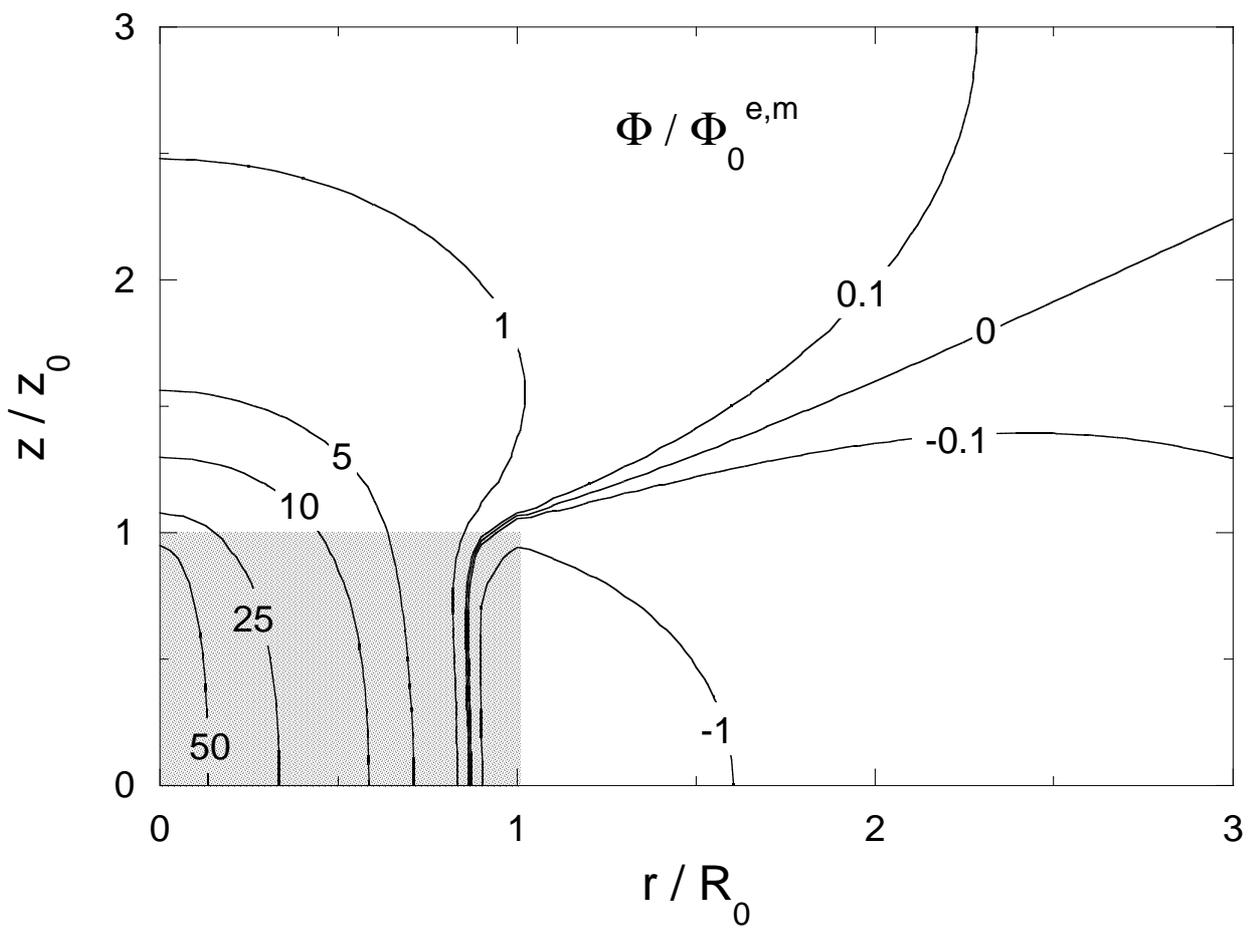
\end{document}